\documentclass[10pt,conference]{IEEEtran}
\IEEEoverridecommandlockouts

\usepackage{cite}
\usepackage{amsmath,amssymb,amsfonts}
\usepackage[T1]{fontenc}
\usepackage{bm, bbm}
\usepackage{pifont}
\usepackage{algorithmic}
\usepackage{graphicx}
\usepackage{textcomp}
\usepackage{xcolor}
\usepackage{physics}
\usepackage{multirow}
\usepackage{enumitem}
\usepackage{subfigure}
\usepackage[linesnumbered,ruled,vlined]{algorithm2e}

\def\BibTeX{{\rm B\kern-.05em{\sc i\kern-.025em b}\kern-.08em
    T\kern-.1667em\lower.7ex\hbox{E}\kern-.125emX}}
\begin{document}

\title{QSeer: A Quantum-Inspired Graph Neural Network for Parameter Initialization in Quantum Approximate Optimization Algorithm Circuits}

\author{
\begin{tabular}{ccc}
Lei Jiang & Chi Zhang & Fan Chen\\
\multicolumn{3}{c}{Indiana University Bloomington}\\
\multicolumn{3}{c}{\{jiang60,czh4,fc7\}\@iu.edu}\\
\end{tabular}
 
}

\maketitle

\begin{abstract}
To mitigate the barren plateau problem, effective parameter initialization is crucial for optimizing the Quantum Approximate Optimization Algorithm (QAOA) in the near-term Noisy Intermediate-Scale Quantum (NISQ) era. Prior physics-driven approaches leveraged the optimal parameter concentration phenomenon, utilizing medium values of previously optimized QAOA parameters stored in databases as initialization for new graphs. However, this medium-value-based strategy lacks generalization capability. Conversely, prior computer-science-based approaches employed graph neural networks (GNNs) trained on previously optimized QAOA parameters to predict initialization values for new graphs. However, these approaches neglect key physics-informed QAOA principles, such as parameter concentration, symmetry, and adiabatic evolution, resulting in suboptimal parameter predictions and limited performance improvements. Furthermore, no existing GNN-based methods support parameter initialization for QAOA circuits with variable depths or for solving weighted Max-Cut problems.

This paper introduces QSeer, a quantum-inspired GNN designed for accurate QAOA parameter prediction. First, we propose a quantum-inspired input data normalization technique to ensure a consistent input scale and mitigate the influence of varying feature magnitudes. By integrating key physics-informed QAOA principles, such as parameter concentration, symmetry, and adiabatic evolution, this approach enhances QSeer’s training stability and convergence. Second, we encode Max-Cut edge weights as edge attributes in QSeer’s GNN framework, enabling parameter prediction for QAOA circuits solving both unweighted and weighted Max-Cut problems. Third, we incorporate the circuit depth $p$ as an input, allowing QSeer to generalize parameter predictions across different QAOA depths. Compared to prior physics- and computer-science-driven methods, QSeer improves the initial approximation ratio and convergence speed of QAOA circuits across diverse graphs by $6\%\sim68\%$ and $5\times\sim10\times$, respectively.
\end{abstract}


\section{Introduction}

The Quantum Approximate Optimization Algorithm (QA-OA)~\cite{Shaydulin:QCS2021,Lee:Mathematics2024,Zentilini:QCE2024,Liang:DAC2024} is a leading variational quantum algorithm (VQA) that employs a parameterized quantum circuit to evaluate a predefined cost function, while a classical optimizer iteratively tunes the circuit parameters. QAOA has demonstrated significant potential in solving combinatorial optimization problems such as Max-Cut~\cite{wang2018quantum}, exhibiting early indications of quantum advantage~\cite{Shaydulin:Science2024} and remaining implementable on near-term Noisy Intermediate-Scale Quantum (NISQ) devices.

Despite its promise, optimizing QAOA circuits remains challenging. The expressive power of QAOA circuits depends on the number of layers~\cite{Lee:Mathematics2024}, with each layer introducing two trainable parameters, $\beta$ and $\gamma$. However, increasing the number of layers, particularly in the presence of many qubits, exacerbates the barren plateau problem~\cite{wang2021noise,zhou2020quantum}, causing gradient magnitudes to vanish exponentially and trapping circuit parameters in local minima. Consequently, effective parameter initialization strategies~\cite{Lee:Mathematics2024,Liang:DAC2024,Shaydulin:QCS2021} are essential for mitigating the impact of barren plateaus.

Research in physics and computer science has independently investigated parameter initialization methods for QAOA, each from its own perspective. However, these efforts remain largely disconnected, preventing any single approach from achieving optimal performance.

Physics-based approaches primarily focus on optimal parameter reuse~\cite{venturelli2024investigating,falla2024graph,Shaydulin:QCS2021,Lee:Mathematics2024,brandao2018fixed} for QAOA circuit initialization, leveraging the ``\textit{optimal parameter concentration}'' phenomenon~\cite{brandao2018fixed,Akshay:PRA2021}, where optimal QAOA parameters exhibit high similarity across different graph instances. Consequently, parameters optimized for smaller graphs can be transferred and reused on larger graphs to approximate solutions, particularly when the smaller graphs are regular. Additionally, certain graphs exhibit inherent symmetries~\cite{lotshaw2021empirical} (e.g., identical vertex groups), which manifest in the QAOA parameter space. These symmetries allow specific angle transformations that yield equivalent computational effects~\cite{Shi:SEC2022}. The optimization of QAOA circuit parameters also aligns closely with the adiabatic quantum computation process~\cite{farhi2000quantum}, where the mixer Hamiltonian is gradually suppressed (i.e., decreasing $\beta$) while the cost Hamiltonian is progressively activated (i.e., increasing $\gamma$). Both parameter symmetry and adiabatic evolution contribute to reducing the effective search space, enhancing the feasibility of parameter transferability. Building on these insights, physics-based studies~\cite{Shaydulin:QCS2021,brandao2018fixed} have constructed databases of optimal QAOA parameters for various small graphs, using the medium parameter values in these databases to initialize circuits for larger graphs. However, this ``medium''-value-based reuse strategy lacks generalization capability, limiting its effectiveness in optimizing QAOA circuits for previously unseen graphs~\cite{Lee:Mathematics2024}. Furthermore, even for graphs stored in the database, variations in circuit configurations, such as the number of layers~\cite{Shaydulin:QCS2021}, can significantly affect performance.

In contrast, computer-science-based approaches typically formulate the selection of initialization parameters for QAOA circuits solving Max-Cut on various graphs as a regression problem. These methods leverage neural networks, including multilayer perceptrons (MLPs)~\cite{li2023proactively,amosy2024iteration} and more advanced graph neural networks (GNNs)~\cite{jain2022graph,Liang:DAC2024}, to model the relationship between graph features and optimal QAOA circuit parameters. However, most existing neural-network-based parameter selection methods are trained on datasets comprising QAOA parameters for Max-Cut on random graphs without explicitly incorporating key structural properties such as the trends of $\beta$ and $\gamma$ across layers, parameter concentration, or angle symmetries. As a result, the parameters predicted by these models~\cite{jain2022graph,Liang:DAC2024,li2023proactively,amosy2024iteration} yield only moderate improvements over random initialization when applied to QAOA circuits solving Max-Cut on diverse graph structures. Additionally, existing neural-network-based approaches~\cite{jain2022graph,Liang:DAC2024,li2023proactively,amosy2024iteration} are typically designed to generate parameters for QAOA circuits with a fixed number of layers. Consequently, they lack the ability to generalize parameter predictions across different circuit depths or capture trends in parameter evolution as the number of layers increases. Furthermore, previous neural network-based techniques do not support parameter initialization for weighted Max-Cut problems. These limitations lead to suboptimal performance when initializing QAOA circuits on new graphs, particularly those with unseen layer configurations or weighted edge structures.

To address these challenges, we propose \textit{QSeer}, a quantum-inspired GNN designed for accurate QAOA circuit parameter initialization, improving both convergence quality and training efficiency. Our contributions are summarized as follows:
\begin{itemize}[leftmargin=*, nosep, topsep=0pt, partopsep=0pt]
\item We introduce a quantum-inspired training data normalization technique that ensures a consistent scale for QAOA circuit parameters by integrating key physics-informed principles, including parameter concentration, symmetry, and adiabatic evolution. This enhances QSeer’s training stability and convergence.

\item We incorporate Max-Cut edge weights as edge attributes within QSeer’s GNN framework, enabling parameter prediction for QAOA circuits solving both unweighted and weighted Max-Cut problems.

\item We encode the circuit depth $p$ as an input, allowing QSeer to generalize parameter predictions across different QAOA circuit depths.

\item Compared to prior physics- and computer science-driven methods, on average, QSeer improves the initial approximation ratio and convergence speed of QAOA circuits across a wide range of graphs by $6\%\sim68\%$ and $5\times\sim10\times$, respectively.
\end{itemize}

\begin{figure*}[t!]
\centering
\includegraphics[width=0.7\linewidth]{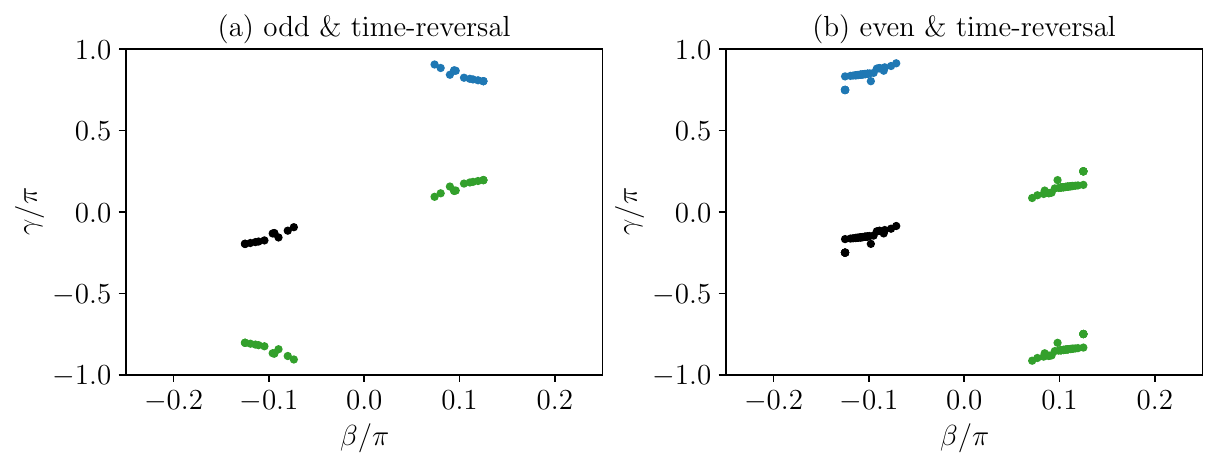}
%\vspace{-0.1in}
\caption{The optimal parameter concentration and symmetry in single-layer ($p=1$) QAOA circuits solving unweighted Max-Cut on regular graphs with degrees ranging from 2 to 8: (a) the time-reversal symmetry on odd-degree regular graphs, and (b) the time-reversal symmetry on even-degree regular graphs.}
\label{f:qaoa_symmetry_all}
\vspace{-0.1in}
\end{figure*}

\begin{figure*}[t!]
\centering
\includegraphics[width=0.7\linewidth]{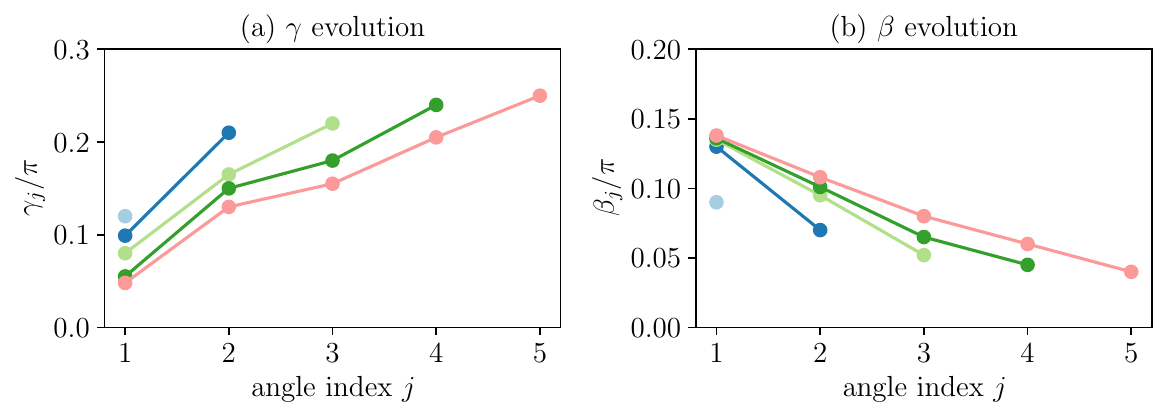}
%\vspace{-0.1in}
\caption{The evolution of optimal $\gamma$ and $\beta$ parameters in QAOA circuits solving unweighted Max-Cut on a 10-node Erd\"os-R\'enyi graph with an edge probability of 0.7, shown as a function of angle index $j$ at a fixed circuit depth $p$.}
\label{f:qaoa_para_index}
\vspace{-0.1in}
\end{figure*}

\section{Background}

\subsection{Quantum Approximate Optimization Algorithm}

The Quantum Approximate Optimization Algorithm (QA-OA)~\cite{Lee:Mathematics2024} aims to maximize the expectation of a cost Hamiltonian $H_Z$ with respect to the circuit state $|\psi(\bm{\gamma}, \bm{\beta})\rangle$, evolved through alternating operators:
\begin{equation}
|\psi_p(\bm{\gamma}, \bm{\beta})\rangle = \prod_{j=1}^{p} e^{-i\beta_j H_x} e^{-i\gamma_j H_z} |+\rangle^{\otimes n},
\label{e:evol_qaoa}
\end{equation}
where $\gamma = (\gamma_1, \gamma_2, \dots, \gamma_p)$ and $\beta = (\beta_1, \beta_2, \dots, \beta_p)$ are the $2p$ variational parameters, with $\bm{\gamma} \in [-\pi, \pi)^p$ and $\bm{\beta} \in [-\pi/2, \pi/2)^p$. The initial state $|+\rangle^{\otimes n}$ represents $n$ qubits in the ground state of the mixer Hamiltonian $H_x = \sum_{j=1}^{n} X_j$, where $X_j$ is the Pauli $X$ operator acting on the $j_{th}$ qubit.

\subsection{QAOA for Max-Cut}

QAOA is particularly effective for combinatorial optimization problems, with one of its primary applications being the NP-complete Max-Cut problem~\cite{guerreschi2019qaoa}. This paper focuses on employing QAOA for solving Max-Cut. The objective of Max-Cut\footnote{Unless explicitly stated otherwise, Max-Cut in this paper refers to the unweighted Max-Cut problem.} is to partition a graph into two subsets while maximizing the number of edges between them. The cost Hamiltonian for a graph $G = (V, E)$ is:
\begin{equation}
H_z = \frac{1}{2} \sum_{(j,k) \in E} w_{j,k} (\mathbbm{1} - Z_j Z_k),
\end{equation}
where $Z_j$ is the Pauli $Z$ operator acting on the $j$-th qubit, and $w_{j,k}$ represents the weight of edge $(j,k)$. When $w_{j,k}=1$ for all $(j,k) \in E$, the problem is unweighted. The average absolute edge weight of a graph is defined as:
\begin{equation}
\overline{|w|}=\frac{1}{|E|}\sum_{(j,k) \in E} |w_{j,k}|.
\end{equation}
The operators $Z_j Z_k$ are applied to qubits $j$ and $k$ for each edge $(j,k)$ in the graph. The expectation value of $H_z$ with respect to the QAOA circuit state from Equation~\ref{e:evol_qaoa} is:
\begin{equation}
F_p(\bm{\gamma}, \bm{\beta}) \equiv \langle \psi_p (\bm{\gamma}, \bm{\beta}) | H_z | \psi_p (\bm{\gamma}, \bm{\beta}) \rangle,
\end{equation}
where $p$ denotes the circuit depth. Solving Max-Cut with QAOA involves minimizing the negative of the expectation value of $H_z$ with respect to the variational parameters $\bm{\gamma}$ and $\bm{\beta}$, achieved via a classical optimizer:
\begin{equation}
(\bm{\gamma}^*, \bm{\beta}^*) \equiv \arg \max_{\bm{\gamma}, \bm{\beta}} F (\bm{\gamma}, \bm{\beta}),
\end{equation}
where the superscript $*$ denotes the optimal parameters. The \textit{approximation ratio} $\alpha$~\cite{Shaydulin:QCS2021} is defined as:
\begin{equation}
\alpha \equiv \frac{F(\bm{\gamma}^*, \bm{\beta}^*)}{C_{\max}},
\end{equation}
where $C_{\max}$ is the maximum cut value of the graph. The approximation ratio $\alpha$ serves as a key performance metric, quantifying the closeness of the QAOA solution to the optimal solution. It satisfies $\alpha \in [0,1]$, where $\alpha = 1$ corresponds to an optimal cut. As the circuit depth $p$ increases, $\alpha$ approaches 1. However, for large $p$, the barren plateau phenomenon arises~\cite{Lee:Mathematics2024}, causing gradient magnitudes to vanish exponentially. Consequently, randomly initialized variational parameters often lead QAOA circuits to become trapped in local minima. To mitigate this issue, effective initialization strategies for variational parameters are essential at the start of the optimization process.

\subsection{Optimal Parameter Concentration in QAOA Circuits}

Optimal parameter concentration is a well-documented property of QAOA circuits, wherein the optimal parameters obtained for solving Max-Cut on small graphs, particularly regular graphs, can be directly reused for larger graphs to approximate solutions without re-optimization~\cite{Akshay:PRA2021,brandao2018fixed}. Formally, optimal parameter concentration implies that for any set of optimal parameters identified for a QAOA circuit with $n$ qubits, there exists at least one parameter set that remains polynomially close (in $n$) and optimal for a QAOA circuit with $n+1$ qubits. As illustrated in Figure~\ref{f:qaoa_symmetry_all}, the variational parameters $\beta$ and $\gamma$ for QAOA circuits solving Max-Cut on different regular graphs exhibit clustering patterns, further reinforcing the phenomenon of optimal parameter concentration.

\subsection{Optimal Parameter Symmetry in QAOA Circuits}
\label{s:opsoc}

As defined in Equation~\ref{e:evol_qaoa}, the initial parameter bounds for the Max-Cut problem are $\bm{\gamma} \in [-\pi, \pi)^p$ and $\bm{\beta} \in [-\pi/2, \pi/2)^p$, reflecting their inherent periodicity~\cite{Lee:Mathematics2024}. However, since the operator $e^{-i (\pi/2) H_x} = X^{\otimes n}$  commutes with the evolution operators in Equation~\ref{e:evol_qaoa}, the periodicity of $\bm{\beta}$ is reduced to $\pi/2$~\cite{zhou2020quantum}. QAOA also exhibits time-reversal symmetry:
\begin{equation}
F_p(\bm{\gamma}, \bm{\beta}) = F_p(-\bm{\gamma}, -\bm{\beta}) = F_p\left( 2\pi - \bm{\gamma}, \frac{\pi}{2} - \bm{\beta} \right).
\label{e:e_fp}
\end{equation}
The second equality follows from the periodicity of $\bm{\gamma}$ and $\bm{\beta}$, with periods of $2\pi$ and $\pi/2$, respectively. From Equation~\ref{e:e_fp}, the landscape of $F_p$ beyond $\gamma=\pi$ is the image of rotation by 180-degree of the landscape within $\gamma=\pi$, corresponding to the reflection of both $\gamma=\pi$ and $\beta=\pi/4$. Exploiting these symmetries, the optimization of $\bm{\gamma}$ and $\bm{\beta}$ can be restricted to the domain $[-\pi/2, \pi/2)^p \times [-\pi/4, \pi/4)^p$, significantly reducing computational redundancy in the search space. Furthermore, additional symmetries arise in regular graphs due to the properties of $e^{-i \pi H_z}$. Specifically, for odd-degree regular graphs, $e^{-i \pi H_z} = Z^{\otimes n}$, while for even-degree regular graphs, $e^{-i \pi H_z} = \mathbbm{1}$. Consequently, for unweighted regular graphs, the optimization bounds of $\bm{\gamma}$ and $\bm{\beta}$ can be further restricted to $[-\pi/4, \pi/4)^p \times [-\pi/4, \pi/4)^p$, further improving computational efficiency. Figure~\ref{f:qaoa_symmetry_all} illustrates the time-reversal symmetry of optimal parameters in QAOA circuits solving unweighted Max-Cut on regular graphs with degrees ranging from 2 to 8.

\subsection{Optimal Parameter Adiabatic Evolution in QAOA Circuits}
\label{s:opae}
The parameter optimization process in QAOA closely follows the adiabatic quantum computation~\cite{farhi2000quantum}, where the variational parameters $\beta$ and $\gamma$ correspond to discrete time steps in an adiabatic evolution. During this process, the mixer Hamiltonian $H_x$ is gradually suppressed (decreasing $\beta$), while the cost Hamiltonian $H_z$ is progressively activated (increasing $\gamma$)~\cite{Cook:QCE2020}. The time-dependent Hamiltonian governing this evolution is given by:
\begin{equation}
H(t) = (1 - t/T) H_x + (t/T) H_z,
\end{equation}
where $T$ denotes the total runtime. The corresponding unitary evolution can be discretized as:
\begin{equation}
e^{-i \int_{0}^{T} H(t) dt} \approx \prod_{j=1}^{p} e^{-i H(j\Delta t) \Delta t},
\label{e:e_adiabatic}
\end{equation}
where $t = \Delta t$. Applying the first-order Lie-Suzuki-Trotter decomposition to Equation~\ref{e:e_adiabatic} yields:
\begin{equation}
e^{-i \int_{0}^{T} H(t) dt} \approx \prod_{j=1}^{p} e^{-i (1 - j\Delta t / T) H_x \Delta t} e^{-i (j\Delta t / T) H_z \Delta t}.
\label{e:e_lst_deco}
\end{equation}
By substituting $\gamma_j = (j\Delta t/T) \Delta t$ and $\beta_j = (1 - j\Delta t/T) \Delta t$, the QAOA form in Equation~\ref{e:evol_qaoa} is recovered. The discretization in Equation~\ref{e:e_adiabatic} partitions the total runtime $T$ into $p$ steps, with $\Delta t = T/p$, leading to the parameter-depth relation:
\begin{equation}
\gamma_j^p = \frac{j}{p} \Delta t; \quad \beta_j^p = \left( 1 - \frac{j}{p} \right) \Delta t.
\end{equation}
Thus, $\gamma_j$ increases with $j$, while $\beta_j$ decreases with $j$. However, due to the approximations in Equations~\ref{e:e_adiabatic} and~\ref{e:e_lst_deco}, neither $\gamma_j$ nor $\beta_j$ is strictly linear with respect to $j$. We illustrate the variation of optimal $\gamma$ and $\beta$ values in QAOA circuits solving unweighted Max-Cut on a 10-node Erd\"os-R\'enyi graph with an edge probability of 0.7.

\section{Related Work}

\textbf{Optimal Parameter Reuse}. Physics-based approaches exploit the optimal parameter concentration phenomenon~\cite{brandao2018fixed,Akshay:PRA2021}, where QAOA parameters exhibit strong similarity across different graphs, enabling parameter transfer from smaller to larger graphs~\cite{venturelli2024investigating,falla2024graph,Shaydulin:QCS2021,Lee:Mathematics2024,brandao2018fixed}. Additionally, parameter symmetries~\cite{lotshaw2021empirical,Shi:SEC2022} and adiabatic evolution paths~\cite{Cook:QCE2020} reduce the search space and improve transferability. Prior studies~\cite{Shaydulin:QCS2021,brandao2018fixed} leveraged these insights to construct databases of optimal QAOA parameters for small graphs, using medium-scale parameter values to initialize circuits for larger graphs. However, this medium-value-based strategy lacks generalization, leading to suboptimal performance on previously unseen graphs~\cite{Lee:Mathematics2024}. Moreover, even for stored graphs, variations in circuit configurations, such as layer count~\cite{Shaydulin:QCS2021}, significantly impact performance.

\textbf{Neural Networks}. Since VQAs can be approximated by tensor networks, classical computing can be used to optimize a VQA before deploying it on NISQ hardware~\cite{rudolph2023synergistic}. Existing neural network-based methods frame QAOA parameter initialization as a regression task, employing MLPs~\cite{li2023proactively,amosy2024iteration} or GNNs~\cite{jain2022graph,Liang:DAC2024} to map graph features to optimal QAOA parameters. However, these methods train on random Max-Cut instances without incorporating QAOA quantum properties such as parameter concentration, symmetry, or adiabatic evolution. Consequently, predicted parameters~\cite{jain2022graph,Liang:DAC2024,li2023proactively,amosy2024iteration} provide only marginal improvements over random initialization. Furthermore, these methods typically generate parameters for QAOA circuits with a fixed number of layers, limiting their ability to generalize across circuit depths~\cite{jain2022graph,Liang:DAC2024,li2023proactively,amosy2024iteration}. Additionally, these techniques do not support weighted Max-Cut, further restricting their applicability. 

\textbf{Trotterized Quantum Annealing}. QAOA circuit parameters can also be initialized using Trotterized quantum annealing~\cite{pelofske2024short,sack2021quantum}. However, this approach necessitates either direct execution on resource-constrained NISQ devices, or computationally expensive classical simulations. Furthermore, prior knowledge from previous QAOA parameter initializations cannot be accumulated or leveraged, requiring the initialization process to be repeated from scratch for each new circuit.

\textbf{Reinforcement and Meta-Learning}. Reinforcement learning~\cite{Wauters:PRR2020,khairy2019reinforcement,garcia2019quantum} and meta-learning~\cite{Chandarana_2023,wilson2021optimizing} have also been explored for optimizing QAOA parameters in Max-Cut problems. However, these methods demand significant NISQ hardware resources for each initialization. Additionally, when initializing parameters for an unseen QAOA circuit, all learning procedures must be retrained, preventing efficient reuse of previously acquired knowledge.

\section{QSeer}

\begin{figure}[t!]
\centering
\includegraphics[width=0.9\linewidth]{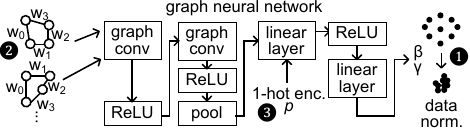}
\vspace{-0.1in}
\caption{An overview of QSeer.}
\label{f:qaoa_graph_all}
\vspace{-0.2in}
\end{figure}

We introduce QSeer, a quantum-inspired GNN designed for efficient QAOA circuit parameter initialization, enhancing both convergence quality and training efficiency for unweighted and weighted Max-Cut problems. An overview of QSeer is presented in Figure~\ref{f:qaoa_graph_all}, where QSeer employs a GNN comprising multiple graph convolutional layers~\cite{wu2019simplifying}, activation layers, pooling layers, and linear layers to process graph inputs and map graph features to QAOA circuit parameters. \ding{182} We propose a quantum-inspired data normalization technique to constrain the distribution of QAOA circuit parameters $\beta$ and $\gamma$ by leveraging key physics-informed principles, including parameter concentration, symmetry, and adiabatic evolution. Similar to classical neural network normalization techniques~\cite{ba2016layer,santurkar2018does}, this approach enhances the training quality of QSeer by stabilizing its gradients, and improves the generalization capability of QSeer by mitigating overfitting. \ding{183} Weighted Max-Cut support is incorporated by encoding Max-Cut edge weights as edge attributes within each graph sample, enabling QSeer to predict QAOA parameters for both unweighted and weighted Max-Cut problems. \ding{184} Circuit depth generalization is achieved by encoding the QAOA circuit depth $p$ as an input to the first linear layer of QSeer, allowing it to adapt parameter predictions across different circuit depths.

\begin{figure*}[t!]
\centering
\subfigure[circuit depth $p=1$.]{
\centering
\includegraphics[width=2in]{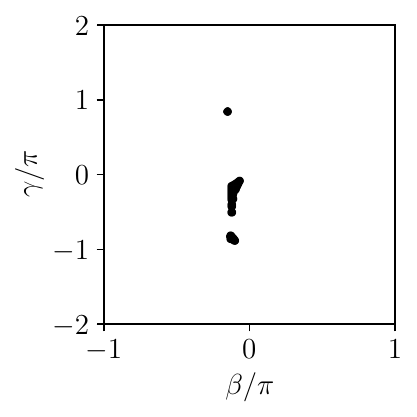}
\label{f:qaoa_unnorm_all1}
}
\hspace{-0.1in}
\subfigure[circuit depth $p=2$.]{
\centering
\includegraphics[width=2in]{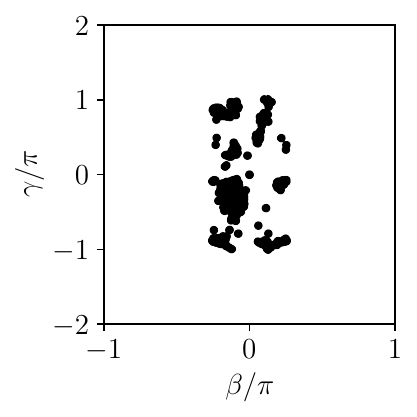}
\label{f:qaoa_unnorm_all2}
}
\hspace{-0.1in}
\subfigure[circuit depth $p=3$.]{
\centering
\includegraphics[width=2in]{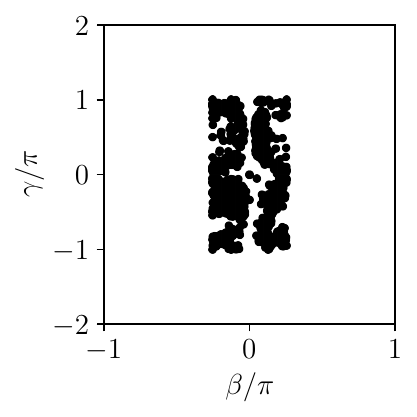}
\label{f:qaoa_unnorm_all3}
}
\vspace{-0.1in}
\caption{The optimal parameter distribution of QAOA circuits solving unweighted Max-Cut without data normalization.}
\label{f:qaoa_unnorm_all}
\vspace{-0.1in}
\end{figure*}

\begin{figure*}[t!]
%\vspace{-0.1in}
\centering
\subfigure[circuit depth $p=1$.]{
\centering
\includegraphics[width=2in]{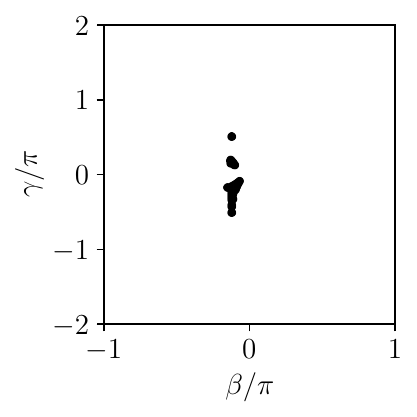}
\label{f:qaoa_normal_all1}
}
\hspace{-0.1in}
\subfigure[circuit depth $p=2$.]{
\centering
\includegraphics[width=2in]{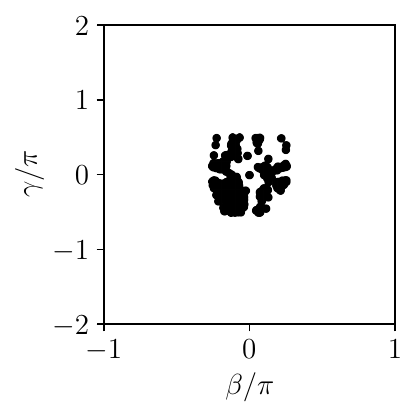}
\label{f:qaoa_normal_all2}
}
\hspace{-0.1in}
\subfigure[circuit depth $p=3$.]{
\centering
\includegraphics[width=2in]{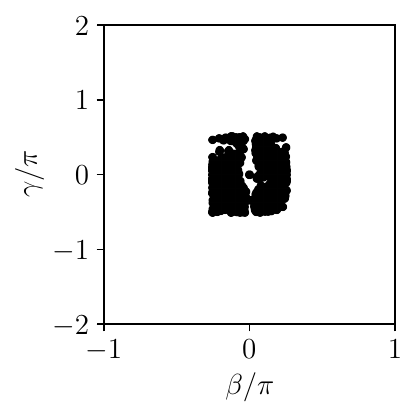}
\label{f:qaoa_normal_all3}
}
\vspace{-0.1in}
\caption{The optimal parameter distribution of QAOA circuits solving unweighted Max-Cut with our quantum-inspired data normalization.}
\label{f:qaoa_normal_all}
\vspace{-0.2in}
\end{figure*}

\subsection{Quantum-Inspired Data Normalization}

To enhance QSeer's learning efficiency on optimal parameters from a limited set of QAOA circuits solving weighted and unweighted Max-Cut, we introduce a quantum-inspired data normalization technique. This method constrains the parameter ranges to $\bm{\gamma} \in [-\pi/2, \pi/2)^p$ and $\bm{\beta} \in [-\pi/4, \pi/4)^p$, where $p$ denotes the circuit depth. Originally, QAOA variational parameters are initialized within $\bm{\gamma} \in [-\pi, \pi)^p$ and $\bm{\beta} \in [-\pi/2, \pi/2)^p$, as shown in Figure~\ref{f:qaoa_unnorm_all}, which depicts the distribution of optimal parameters across approximately 800K QAOA circuits solving unweighted Max-Cut. Regardless of circuit depth, all values of $\bm{\gamma}$ and $\bm{\beta}$ remain within these bounds. By leveraging the symmetry properties of QAOA introduced in Section~\ref{s:opsoc}, we normalize the variational parameters to the reduced range of $[-\pi/2, \pi/2)^p \times [-\pi/4, \pi/4)^p$. Figure~\ref{f:qaoa_normal_all} illustrates the distribution of normalized parameters, demonstrating a significant reduction in their range. This improved concentration simplifies QSeer's ability to predict parameters for QAOA circuits solving Max-Cut on unseen graphs. Additionally, we impose the adiabatic evolution trend of QAOA parameters, as introduced in Section~\ref{s:opae}, by enforcing that $\bm{\gamma}$ increases with $p$ while $\bm{\beta}$ decreases with $p$. This further reduces the learning complexity for QSeer, improving parameter generalization. Despite the change in parameter ranges, our normalization technique preserves the original loss values of QAOA circuits, as it inherently accounts for the quantum properties of QAOA parameterization.

\subsection{Support for Weighted Max-Cut}

Unlike previous network-based QAOA parameter predictors~\cite{Liang:DAC2024,Zentilini:QCE2024,jain2022graph,falla2024graph,amosy2024iteration}, QSeer is capable of generating initialization parameters for QAOA circuits solving weighted Max-Cut. A GNN~\cite{kipf2016semi,wu2020comprehensive} inductively learns node representations by recursively aggregating and transforming feature vectors from neighboring nodes. The update of a graph convolution layer~\cite{kipf2016semi} consists of three fundamental operations: message passing, message aggregation, and node representation updating. The message passing is defined as:
\begin{equation}
\mathbf{m}_{vu}^{(l)} = \text{MSG}(\mathbf{h}_u^{(l-1)}, \mathbf{h}_v^{(l-1)}, e_{vu}),
\label{e:m_pass}
\end{equation}
where $\mathbf{h}_u^{(l-1)}$ represents the embedded state of node $u$ at layer $l-1$, and $e_{vu}$ denotes the weighted edge between nodes $v$ and $u$. The message aggregation step is given by:
\begin{equation}
\mathbf{M}_i^{(l)} = \text{AGG}(\{\mathbf{m}_{vu}^{(l)}, e_{vu} \mid v \in \mathcal{N}(u)\}),
\label{e:m_agg}
\end{equation}
where $\mathcal{N}(u)$ represents the neighborhood of node $u$, from which information is collected to update its aggregated message $\mathbf{M}_i$. The term $\mathbf{m}_{vu}^{(l)}$ represents the message passed from node $v$ to node $u$ at layer $l$. Finally, the node representation update step is performed as:
\begin{equation}
\mathbf{h}_u^{(l)} = \text{UPDATE}(\mathbf{M}_u^{(l)}, \mathbf{h}_u^{(l-1)}).
\end{equation}
By incorporating edge weights $e_{vu}$ in Equations~\ref{e:m_pass} and~\ref{e:m_agg}, QSeer effectively integrates weighted edges into its GNN-based parameter prediction framework, enabling support for QAOA circuits solving weighted Max-Cut problems. However, most GNNs are not inherently designed to process negative edge weights~\cite{kipf2016semi,wu2020comprehensive}. To address this, we apply a normalization technique that maps all edge weights to the range $[0,1]$:
\begin{equation}
W_{e_{vu}}^* = \frac{W_{e_{vu}} - W_{min}}{W_{max}- W_{min}},
\label{e:normal_edge}
\end{equation}
where $W_{e_{vu}}$ is the original edge weight, $W_{e_{vu}}^*$ is the normalized edge weight, $W_{max}$ represents the maximum edge weight, and $W_{min}$ denotes the minimum edge weight in the graph. Equation~\ref{e:normal_edge} ensures numerical stability and compatibility with standard GNN architectures while preserving relative weight differences.

\begin{figure}[t!]
\centering
\includegraphics[width=0.6\linewidth]{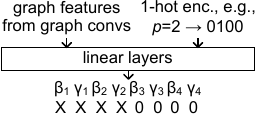}
\vspace{-0.1in}
\caption{The circuit depth $p$ is an input to the first linear layer.}
\label{f:qaoa_graph_one}
\vspace{-0.2in}
\end{figure}

\subsection{Support for Variable Circuit Depths}

Existing network-based QAOA parameter predictors~\cite{Liang:DAC2024,Zentilini:QCE2024,jain2022graph,falla2024graph,amosy2024iteration} are limited to predicting parameters for QAOA circuits with a fixed depth ($p$). To enable parameter prediction across different circuit depths, QSeer incorporates $p$ as an input to the first linear layer of its GNN architecture, as illustrated in Figure~\ref{f:qaoa_graph_one}. The circuit depth $p$ is encoded using one-hot representation and concatenated with the processed graph features before being fed into the linear layer. QSeer supports a maximum circuit depth (e.g., $p=4$), allowing it to generate up to $2p$ (e.g., 8) variational parameters. When the input depth $p$ is smaller than the maximum supported value (e.g., $p=2$), only the first $2p$ (e.g., 4) generated parameters are utilized, while the remaining parameters (e.g., 4) are set to zero. This design ensures flexibility in adapting QAOA parameter predictions to circuits of varying depths.

\begin{figure*}[t!]
%\vspace{-0.1in}
\centering
\subfigure[training data.]{
\centering
\includegraphics[width=3in]{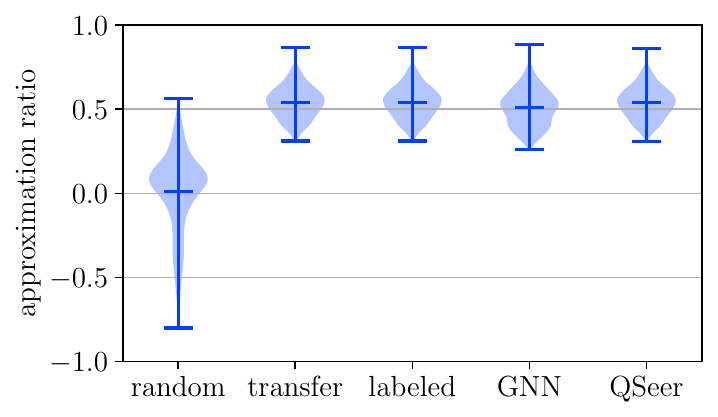}
\label{f:qaoa_train_result}
}
\hspace{-0.1in}
\subfigure[testing data.]{
\centering
\includegraphics[width=3in]{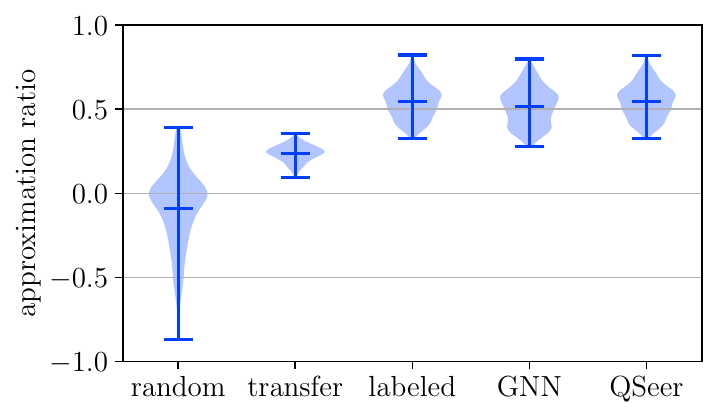}
\label{f:qaoa_test_result}
}
\vspace{-0.1in}
\caption{The parameter initialization of QAOA circuits solving unweighted Max-Cut with quantum-inspired data normalization.}
\label{f:qaoa_unweighted_all}
\vspace{-0.1in}
\end{figure*}

\begin{figure*}[t!]
%\vspace{-0.1in}
\centering
\subfigure[training data.]{
\centering
\includegraphics[width=3in]{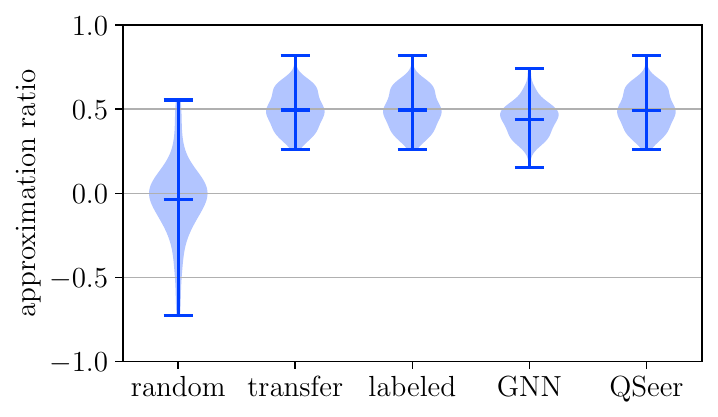}
\label{f:qaoa_train_result0}
}
\hspace{-0.1in}
\subfigure[testing data.]{
\centering
\includegraphics[width=3in]{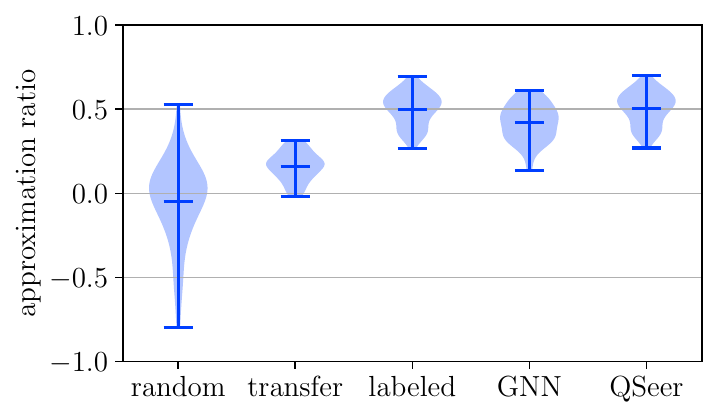}
\label{f:qaoa_test_result0}
}
\vspace{-0.1in}
\caption{The parameter initialization of QAOA circuits solving weighted Max-Cut with quantum-inspired data normalization.}
\label{f:qaoa_weighted_all}
\vspace{-0.1in}
\end{figure*}

\begin{figure*}[t!]
%\vspace{-0.1in}
\centering
\subfigure[unweighted graphs.]{
\centering
\includegraphics[width=3in]{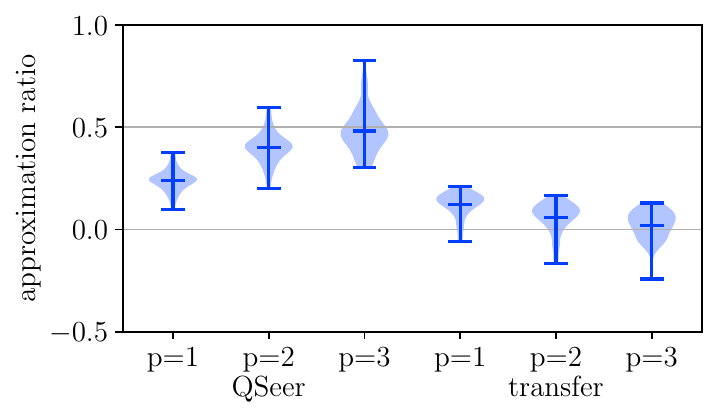}
\label{f:qaoa_unseen_now_result}
}
\hspace{-0.1in}
\subfigure[weighted graphs.]{
\centering
\includegraphics[width=3in]{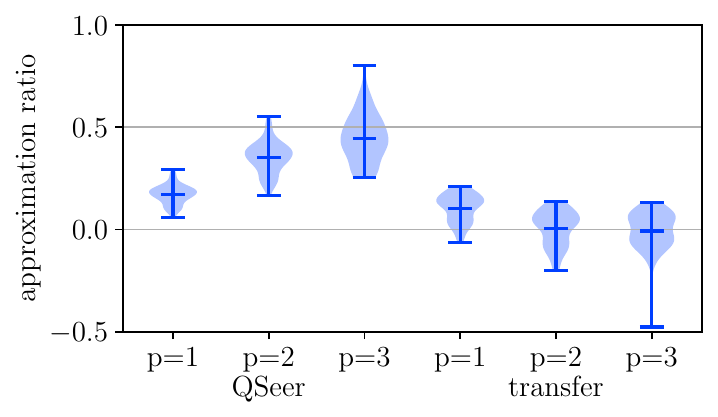}
\label{f:qaoa_unseen_w_result}
}
\vspace{-0.1in}
\caption{The parameter initialization of QAOA circuits solving unweighted and weighted Max-Cut on unseen graphs.}
\label{f:qaoa_unseen_result}
\vspace{-0.2in}
\end{figure*}

\section{Experimental Methodology}

\textbf{Dataset}. For unweighted Max-Cut, we use the dataset from~\cite{lotshaw2021empirical}, which contains all connected non-isomorphic graphs of size $n \leq 9$, totaling approximately 800K graphs. Each graph includes QAOA cost values and optimal parameters for depths $p \leq 3$, obtained from the best results of 50 ($p=1$), 100 ($p=2$), and 1000 ($p=3$) calls to the Goemans-Williamson optimizer~\cite{patti2023quantum} with random angle seeds. The results for $p=1$ were validated against exact methods. For weighted Max-Cut, we use the dataset from~\cite{shaydulin2023parameter}, consisting of 34,701 weighted graphs with $n \leq 20$. This dataset includes all non-isomorphic 8-node graphs, 300 Erd\"os-R\'enyi random graphs with 14 nodes, and 50 Erd\"os-R\'enyi random graphs with 20 nodes, where edge probabilities are set to 0.5. Edge weights follow uniform and exponential distributions. QAOA cost values and optimal parameters for $p \leq 3$ were computed using the best results from 50 ($p=1$), 100 ($p=2$), and 1000 ($p=3$) calls to the Goemans-Williamson optimizer with random angle seeds. We randomly split all data into training, validation, and testing sets in an $8:1:1$ ratio. Additionally, we construct an unseen graph dataset comprising 5K regular graphs and 5K Erd\"os-R\'enyi graphs, both with 10 to 12 nodes. The edge probabilities for Erd\"os-R\'enyi graphs are randomly sampled from $[0.1, 0.9]$, and edge weights for all graphs are assigned following uniform and exponential distributions.

\textbf{QAOA}. We employ the \texttt{TensorCircuit} simulator~\cite{zhang2023tensorcircuit} to compute the loss of QAOA circuits solving unweighted and weighted Max-Cut across various graphs. \texttt{TensorCircuit} simulates quantum circuits using tensor network methods, enabling efficient evaluation. All QAOA circuits are trained using the \texttt{Adam} optimizer with a learning rate of 0.01. Training and evaluation experiments are conducted on an NVIDIA A100 GPU.

\textbf{GNN}. QSeer is designed with a GNN architecture consisting of one \texttt{GCNConv} layer, two \texttt{GATConv} layers, and a two-layer MLP. Each layer is followed by a \texttt{ReLU} activation function. The \texttt{GCNConv} and \texttt{GATConv} layers process node features and edge weights, with a hidden size of 256. The MLP has a hidden size of $256 + oh$, where $oh$ represents the one-hot encoding length of the circuit depth $p$. QSeer is trained using the \texttt{Adam} optimizer and an \texttt{MSELoss} function for 20 epochs, with a linearly decaying learning rate initialized at 0.01. All training and evaluation experiments are conducted on an NVIDIA A100 GPU.

\textbf{Schemes}. We implement and compare the following parameter initialization techniques for QAOA circuits solving unweighted and weighted Max-Cut problems:
\begin{itemize}[leftmargin=*, nosep, topsep=0pt, partopsep=0pt]
\item \textit{Random}: $\bm{\gamma}$ is randomly initialized in $[-\pi, \pi)^p$, and $\bm{\beta}$ in $[-\pi/2, \pi/2)^p$.
\item \textit{Transfer}: Optimal parameters from existing QAOA data-sets~\cite{Shaydulin:QCS2021,brandao2018fixed} are stored in a database. For unseen unweighted graphs, median values of related QAOA circuit parameters are transferred and used for initialization. Weighted graphs require equation-based adjustments~\cite{shaydulin2023parameter}.
\item \textit{Labeled}: The ground truth optimal parameters of QAOA circuits solving Max-Cut on unweighted and weighted graphs.
\item \textit{GNN}: A GNN-based parameter predictor~\cite{jain2022graph,Liang:DAC2024} trained on our dataset without incorporating data normalization, edge weight support, or circuit depth generalization.
\item \textit{QSeer}: Our proposed GNN predictor trained on our dataset and enhanced with data normalization, edge weight support, and circuit depth generalization.
\end{itemize}

\section{Evaluation and Results}

\subsection{Unweighted Graphs}

We analyze the approximation ratio (AR) results achieved by different parameter initialization schemes for unweighted Max-Cut, as shown in Figure~\ref{f:qaoa_unweighted_all}. For QAOA circuits in the training dataset (Figure~\ref{f:qaoa_train_result}), the \texttt{random} initialization yields an average AR of 0.013. The \texttt{transfer} method retrieves and applies identical parameter values as the \texttt{labeled} scheme, achieving equivalent AR performance. However, due to the large parameter search space, the conventional \texttt{GNN} reduces the average AR by 6.1\%. In contrast, \texttt{QSeer} achieves significantly better generalization than \texttt{GNN}, with only a 0.04\% reduction in AR compared to the \texttt{labeled} scheme. For QAOA circuits in the testing dataset (Figure~\ref{f:qaoa_test_result}), the \texttt{random} initialization continues to yield an average AR of -0.089. Meanwhile, the \texttt{transfer} method, which transfers fixed median parameter values from the training dataset, suffers a 56.7\% reduction in average AR. The weak adaptability of the \texttt{transfer} method limits its best-case, worst-case, and average AR values. Compared to the training dataset, the conventional \texttt{GNN} experiences a 5.9\% AR degradation relative to the \texttt{labeled} scheme. In contrast, \texttt{QSeer} improves the average AR by 0.2\% over the \texttt{labeled} scheme, demonstrating superior parameter generalization than the conventional \texttt{GNN}.

\subsection{Weighted Graphs}

The AR results for different parameter initialization schemes on weighted Max-Cut are shown in Figure~\ref{f:qaoa_weighted_all}. For QAOA circuits solving weighted Max-Cut in the training dataset (Figure~\ref{f:qaoa_train_result0}), the \texttt{random} initialization yields an average AR of -0.034. The AR results of all other schemes are generally lower than their unweighted counterparts, reflecting the increased complexity of cutting weighted graphs. The \texttt{transfer} method and the \texttt{labeled} scheme use the same parameter values within the training dataset, achieving equivalent AR performance. Due to the large variational parameter range, the conventional \texttt{GNN} suffers a 11\% reduction in average AR, while \texttt{QSeer}, leveraging quantum-inspired data normalization, reduces this degradation to only 0.1\% compared to the \texttt{labeled} scheme. For QAOA circuits solving weighted Max-Cut in the testing dataset (Figure~\ref{f:qaoa_test_result0}), the \texttt{random} initialization continues to yield an average AR of -0.046. Compared to the \texttt{labeled} scheme, the \texttt{transfer} method reduces the average AR result by 67.9\%, highlighting the limitations of reusing fixed median parameter values for initializing QAOA circuits on weighted graphs. The conventional \texttt{GNN} exhibits a 15.8\% average AR loss and a 23.5\% worst-case AR loss relative to the \texttt{labeled} scheme, reflecting instability in its predicted initialization parameters. In contrast, \texttt{QSeer} marginally improves the average AR by 0.1\% compared to the \texttt{labeled} scheme, demonstrating its robust generalization capabilities.

\subsection{Support for Various Circuit Depths}

Figure~\ref{f:qaoa_unseen_result} presents the AR results for QAOA circuits with different circuit depths ($p$). For Max-Cut on unseen unweighted graphs (Figure~\ref{f:qaoa_unseen_now_result}), \texttt{QSeer} demonstrates a consistent improvement in minimal, average, and maximum AR as $p$ increases. Specifically, the average AR improves by 67\% when increasing from $p=1$ to $p=2$, and by an additional 21\% when increasing to $p=3$. In contrast, the \texttt{transfer} method, which relies on fixed median parameter transfer, does not guarantee improved AR with increasing $p$. Instead, parameter mismatches accumulate as $p$ increases, leading to a decline in average AR—dropping by 58\% from $p=1$ to $p=2$, and by a further 67\% from $p=2$ to $p=3$. For Max-Cut on unseen weighted graphs (Figure~\ref{f:qaoa_unseen_w_result}), all \texttt{QSeer} schemes experience at least an 8\% reduction in average AR compared to their performance on unweighted graphs, reflecting the increased complexity of weighted Max-Cut. Despite this, \texttt{QSeer} continues to achieve increasing average AR with larger $p$, yielding a 105\% improvement from $p=1$ to $p=2$, and a further 27\% increase from $p=2$ to $p=3$. Conversely, the \texttt{transfer} method struggles with parameter generalization as $p$ increases, with average AR dropping by 95\% from $p=1$ to $p=2$, and by 46\% from $p=2$ to $p=3$. These results highlight the robustness of \texttt{QSeer} in effectively adapting to different circuit depths.

\begin{figure}[t!]
\centering
\subfigure[unweighted graphs.]{
\centering
\includegraphics[width=1.65in]{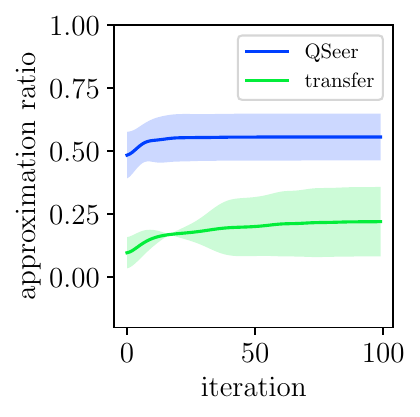}
\label{f:qaoa_unweighted_unseen_training}
}
\hspace{-0.1in}
\subfigure[weighted graphs.]{
\centering
\includegraphics[width=1.65in]{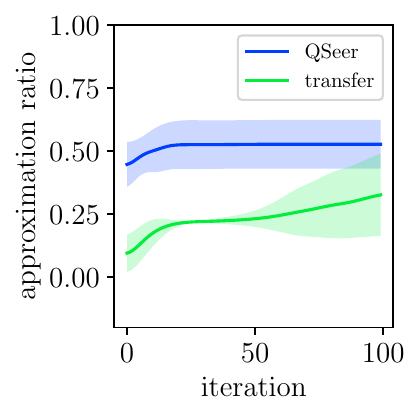}
\label{f:qaoa_weighted_unseen_training}
}
\vspace{-0.1in}
\caption{The AR value changes during the optimizations of QAOA solving Max-Cut on unweighted and weighted graphs.}
\label{f:qaoa_unseen_training}
\vspace{-0.1in}
\end{figure}

\subsection{Final AR Improvement}

We evaluate the final AR improvement of QAOA circuits solving unweighted and weighted Max-Cut on the unseen graph dataset, as presented in Figure~\ref{f:qaoa_unseen_training}. We compare \texttt{QSeer} with the \texttt{transfer} method by initializing QAOA circuits and optimizing them for 100 iterations using the Adam optimizer with a learning rate of 0.01. For unweighted graphs (Figure~\ref{f:qaoa_unweighted_unseen_training}), \texttt{QSeer} stabilizes the final AR within approximately 10 iterations, whereas the \texttt{transfer} method requires $80\sim90$ iterations for stabilization. Additionally, \texttt{QSeer} improves the final AR by 161\% over the \texttt{transfer} method while exhibiting reduced variation in the final AR values. For weighted graphs (Figure~\ref{f:qaoa_weighted_unseen_training}), \texttt{QSeer} stabilizes the final AR within 25 iterations, whereas the \texttt{transfer} method continues to improve AR even after 100 iterations, indicating a prolonged optimization process. After 100 iterations, \texttt{QSeer} outperforms the \texttt{transfer} method by increasing the final AR by 57\%, while maintaining significantly lower variation around the final AR values. These results demonstrate that \texttt{QSeer} not only accelerates QAOA convergence but also achieves higher and more stable final AR values compared to the state-of-the-art \texttt{transfer}-based parameter initialization technique.

\section{Conclusion}
Effective parameter initialization is crucial for mitigating the barren plateau problem and improving QAOA performance in the NISQ era. While prior physics-driven approaches utilize optimal parameter concentration for initialization, they lack generalization to unseen graphs. Conversely, existing GNN-based methods, trained on previously optimized QAOA parameters, fail to incorporate key physics-informed principles, leading to suboptimal predictions. Additionally, no prior approaches support variable circuit depths or weighted Max-Cut problems. To address these limitations, we introduce QSeer, a quantum-inspired GNN designed for accurate QAOA parameter prediction. QSeer integrates a quantum-inspired data normalization technique that constrains parameter distributions based on physics-informed principles, improving training stability and generalization. Furthermore, it encodes Max-Cut edge weights as edge attributes, enabling support for weighted Max-Cut instances. By incorporating circuit depth $p$ as an additional input, QSeer generalizes parameter predictions across varying QAOA depths.

\bibliographystyle{IEEEtranS}
\bibliography{quantum}

\end{document}